# BEAM DIAGNOSTICS SYSTEMS FOR THE NATIONAL IGNITION FACILITY

R. D. Demaret, R. D. Boyd, E. S. Bliss, A. J. Gates, J. R. Severyn
LLNL, Livermore, CA 94550, USA

## Abstract

The National Ignition Facility (NIF) laser focuses 1.8 megajoules of ultraviolet light (wavelength 351 nanometers) from 192 beams into a 600-micrometer-diameter volume. Effective use of this output in target experiments requires that the power output from all of the beams match within 8% over their entire 20-nanosecond waveform. The scope of NIF beam diagnostics systems necessary to accomplish this task is unprecedented for laser facilities. Each beamline contains 110 major optical components distributed over a 510-meter path, and diagnostic tolerances for beam measurement are demanding. Total laser pulse energy is measured with 2.8% precision, and the interbeam temporal variation of pulse power is measured with 4% precision. These measurement goals are achieved through use of approximately 160 sensor packages that measure the energy at five locations and power at three locations along each beamline using 335 photodiodes, 215 calorimeters, and 36 digitizers. Successful operation of such a system requires a high level of automation of the widely distributed sensors. Computer control systems provide the basis for operating the shot diagnostics with repeatable accuracy, assisted by operators who oversee system activities and setup, respond to performance exceptions, and complete calibration and maintenance tasks.

## 1 INTRODUCTION

For the NIF to achieve its performance goal of 1.8 Mega-joules of ultraviolet light on target requires that each beam path is carefully controlled and accurately characterized. Each beam will be operated at up to 80% of its fluence damage threshold and still need to meet the System Design Requirement (SDR) of ≤8% rms deviation in the power delivered. This places stringent derived requirements on the energy, power temporal shape, and fluence measurement systems deployed in the NIF. Table 1 lists the principal requirements for measurements of pulse energy and power, and monitoring of the spatial profile of beam fluence.

Table 1: Tolerances for key diagnostics tasks

| Beam Diagnostics Measurements Requirements | |
|---|---|
| Measure pulse power of all beams (power balance) | The rms deviation less than 8% of the specified power averaged over any 2-ns time interval [1]. |
| Measure pulse energy at 1.053 and 0.351µm | 2.8% |
| Measure pulse temporal shape versus time | 2.8% with ≤ 450psec rise time |
| Record the spatial profile of beam fluence | 2% fluence resolution, 1/125 of beam spatial resolution |

Sensor packages and diagnostics sensors complete the principal diagnostics hardware. Supporting front-end processors (FEP) and diagnostics electronics capture and process the data. The Integrated Computer Controls System (ICCS) [2] will provide the basis for completing shot preparations with repeatable accuracy on a timely basis.

These beam diagnostic functions are accomplished with optical-mechanical and electronic components distributed along each beamline. Figure 1 identifies these components and illustrates the fact that the beam control systems have interfaces with every part of the laser.

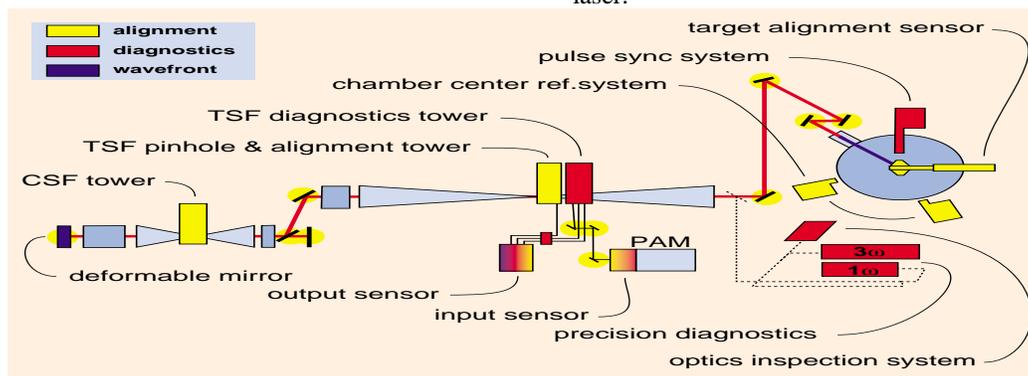

Figure 1: Optical, electronic, and mechanical components distributed along each beam perform beam diagnostic functions. Towers hold components for laser diagnostics functions on eight beams. In the figure, only one beamline is shown for clarity.

## 2 OPTICAL DIAGNOSTICS SYSTEMS

### 2.1 Input Sensor Package (ISP)

The input sensor package is located at the beginning of the laser chain and at the output of the preamplifier module (PAM), Figure 2. A charge coupled device (CCD) camera system is used to capture a nearfield or farfield image of the pulsed beam during a shot. A temporal sample of the beam is collected and launched into a fiberoptic cable to the power diagnostics system. The energy in the beamline is measured using an integrating sphere and collected by the energy diagnostic system. This energy diagnostic is calibrated against an insertable whole beam energy calorimeter.

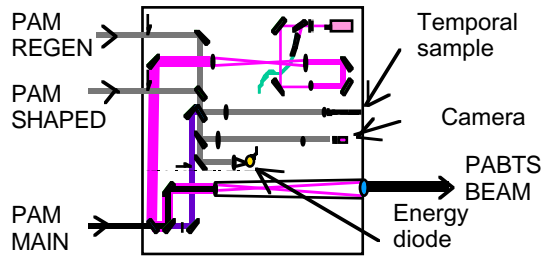

Figure 2: The ISP schematic shows how the PAM beams are sampled for alignment and diagnostics.

### 2.2 Output Sensor Package (OSP)

The output sensor is located below the transport spatial filter at the end of a set of relay optics as shown in Figure 3. This sensor provides diagnostic functions for the full beamline performance of a pair of beams. The diagnostic tasks performed by the OSP completely characterize the output of the main amplifier, measuring wavefront, and recording spatial and temporal pulse shapes. Multiplexing is utilized to record the performance at a reasonable cost. Two nearfield images are fit onto the $1\omega$ CCD camera, so every beam is imaged every shot.

Energy measurement is accomplished for each beam in the relay optics, where a beam sample is directed onto a photodiode assembly. A temporal sample of one beam from every pair of $1\omega$ beams is sent to the power diagnostic as described in the Input Sensor Package (ISP). The CCD cameras image the target plane by collecting light reflecting off of the final focus lens. Full-aperture roving calorimeters, which consist of an array of eight calorimeters that can move to intercept from 1 to 8 of the beams in a bundle, are located in the switchyard and complete the list of diagnostics. These calorimeters are used to calibrate the main amplifier photodiodes.

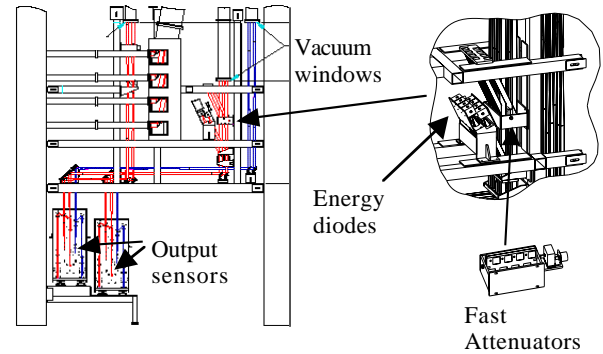

Figure 3: A model of the area under the transport spatial filter. The main beam samples are relayed to the output sensors. There is one energy diode for each beamline sample.

### 2.3 $3\omega$ Diagnostic

The $3\omega$ diagnostics package is located at the target chamber following the frequency conversion crystals. The package measures the energy in the $3\omega$ output beam with a calorimeter and the temporal waveform by directing a beam sample into a $3\omega$ fiberoptic to the power diagnostics system as described in the ISP, see Figure 4.

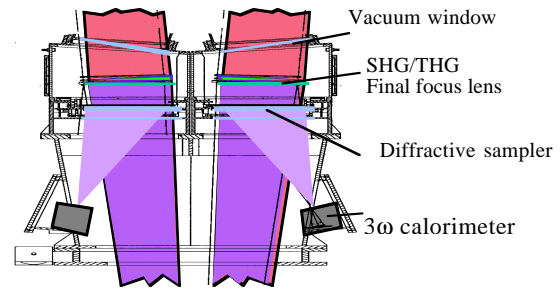

Figure 4: The absorbing glass calorimeter in the FOA gets a sample of the $3\omega$ energy from an off-axis diffractive splitter.

## 3 DIAGNOSTICS MEASUREMENT SYSTEMS

### 3.1 Energy Diagnostics

The Energy Diagnostics system measures the laser pulsed energy to 2.8% accuracy using photodiodes and calorimeters at 432 locations distributed throughout the facility. The measurement electronics need to be accurate to better than 1.0% over a dynamic range of at least $10^3$. An additional 24 calorimeters are used to calibrate the ISP and OSP photodiodes online, minimizing measurement errors induced by diagnostic component changes. These widely distributed sensors are interfaced using an Echelon™ LonTalk™ field bus-based embedded controller architecture to control the energy nodes located at the sensors. Centrally located VME-based PowerPC FEPs use a commercial off-the-shelf (COTS) VME-based Echelon™ LonTalk™ interface to communicate with the remote nodes. To simplify the design, the sensor power is also distributed on the network cable. The energy nodes are designed to minimize measurement errors and include maintenance functions to calibrate and detect energy node failures. Each LLNL-designed node has a signal source used to calibrate and test the node hardware in-situ. All calibration data, both system and hardware specific, is stored on the node in nonvolatile memory to ensure the calibration information remains with the node. Once the node is online and data is collected, the calibration information is used to calculate the absolute energy measured at this location.

*3.2 Power Diagnostics*

The temporal power measurement system (Power Diagnostics or PD) measures the pulse temporal shape to 2.8% accuracy using digital real-time oscilloscopes (digitizers) at 240 locations distributed throughout the facility. Meeting the measurement accuracy requires a digitizer with a dynamic range ≥3100, a record length of 22 ns, and a rise-time of ≤450 ps. In addition, an in-situ calibration system is used to periodically measure system performance for correction of the waveforms. The beam samples are collected from the ISP, OSP, and 3ω diagnostics and transported to centrally located Power Sensor Packages (PSP). Optically delaying each signal sequentially by 50 ns using 8-80 meter cables and converting them into electrical signals using a vacuum photodiode in the PSP multiplexes these signals. The electrical signals are input into a 2-to-1 transformer and recorded on a digitizer's four channels operated at different gains, allowing "overlaying" of the channels to extend the dynamic range of the measurement. These samples are then sequentially recorded on the digitizers to reduce the cost of the system. The 36 COTS digitizers are interfaced to the ICCS using 24 rack-mounted UltraSPARC processors and COTS Ethernet based GPIB interfaces. After the FEP reads the captured channel waveforms, they are corrected using digital signal processing (DSP) algorithms to correct amplitude distortion and normalize their bandwidth. Once the "raw" channel data is corrected, the four channel segments that correspond to a single "beamline" waveform are reconstructed using DSP algorithms to correct for the channel timing skew. This beamline data is then transferred from the FEP, and all beamline waveforms are analyzed offline with the ener-gy data to verify the NIF power balance for the shot.

*3.3 Image Diagnostics*

The measurement of the spatial uniformity of energy distribution within the beam is necessary to operate the facility consistently at 80% of the optical components damage threshold. This requires a 2% fluence (energy per unit area) measurement resolution at 1/125 of the beam spatial resolution. The principal means for monitoring the spatial profile of the beams is to capture images on CCD cameras located in the ISP and OSP described above. These images must then be digitized and processed through the ICCS Video FEP to determine the extent and locations of significant fluence modulation. The video FEP is a rack-mounted UltraSPARC processor and uses six four-channel COTS framegrabbers to capture the video images. All CCD cameras in the system are provided power and the synchronization signal by a Camera Interface Unit (CIU) located in proximity to the camera. The RS-170 video signal from the camera is returned to the CIU and digitized by a framegrabber within the FEP. The interconnection between a CIU and the FEP is either a coaxial cable or multi-mode fiber depending on the specific camera location within the NIF.

## 4 SUMMARY

The flow-down measurement accuracy for diagnosing the NIF created challenging design requirements for these beam diagnostics systems. Through a combination of COTS and LLNL-developed distributed systems combined with in-situ calibration and digital correction, we were able to accomplish these goals.